\RequirePackage[2020-02-02]{latexrelease}
\documentclass[twocolumn,prl,amsmath,amssymb,showpacs,superscriptaddress,floatfix]{revtex4}
\usepackage{graphicx}% Include figure files
\usepackage{bm}% bold math
\usepackage{lineno,hyperref}
\usepackage{epsf}
\usepackage{xcolor}
\begin{document}
%\draft
\title{A rare gas mixture: From rigid to gas-like fluid by a mutual concentration change}%:
%First-Order versus Continuous Transition }

\author{Yu. D. Fomin \footnote{Corresponding author: fomin314@mail.ru}}
\affiliation{Vereshchagin Institute of High Pressure Physics,
Russian Academy of Sciences, Kaluzhskoe shosse, 14, Troitsk,
Moscow, 108840, Russia }

\author{E. N. Tsiok }
\affiliation{Vereshchagin Institute of High Pressure Physics,
Russian Academy of Sciences, Kaluzhskoe shosse, 14, Troitsk,
Moscow, 108840, Russia }

\author{V. N. Ryzhov }
\affiliation{Vereshchagin Institute of High Pressure Physics,
Russian Academy of Sciences, Kaluzhskoe shosse, 14, Troitsk,
Moscow, 108840, Russia }
\date{\today}

\author{V. V. Brazhkin}
\affiliation{Vereshchagin Institute of High Pressure Physics,
Russian Academy of Sciences, Kaluzhskoe shosse, 14, Troitsk,
Moscow, 108840, Russia }
\date{\today}

\begin{abstract}

For a number of mixtures of rare gases at high pressures, sound
speed minima are experimentally observed depending on the
concentration. This behavior has not yet been explained. We have
studied the behavior of a mixture of argon and helium using
computer simulation. Sound speed minima have been discovered at a
certain concentration, which is in good agreement with
experimental data. It is shown that this behavior is due to the
fact that the P and T parameters for gas mixtures are near the
Frenkel line, separating the states of “rigid” and quasi-gas
fluid.

\end{abstract}


\maketitle

%\begin{keywords}
%  boiling line, near-critical maxima (Widom line), phase equilibria of mixtures
%\end{keywords}

\section{Introduction}

Speed of sound $c_s$ is an important thermodynamic characteristic
of a substance, which is relatively easily measurable
experimentally. It allows obtaining other thermodynamic
properties, first of all an equation of state, i.e. the density as
a function of pressure at given temperature. A speed of sound also
demonstrates different behavior in a dense liquid and in a gas:
while in the former case it decreases upon isobaric heating, in
the latter it increases. Some other properties of fluid also
demonstrate qualitatively different behavior in liquid and gas
states, for instance, shear viscosity. As a result, it also
demonstrates a minimum. In our recent works a special line
separating liquid-like and gas-like regimes of fluid (the Frenkel
line) was proposed \cite{ufn-fr,pre-fr,jepc-fr,scf-fr}. It is
remarkable that the minima of all quantities with different
behavior in liquid and gas states are close to the Frenkel line.

Although the speeds of sound of pure fluids are sufficiently well
documented, much less is known about the speed of sound in
mixtures, especially at elevated pressures. Three types of
concentration dependence of the speed of sound in mixtures are
reported \cite{73}: (i) with a minimum of $c_s$ as a function of
the component concentration; (ii) with a maximum and (iii) with a
bend. However, up to now there is no firm theoretical
understanding of the reasons for different behavior of $c_s$ in
mixtures.

The minimum of concentration dependence of the speed of sound is
reported even in very simple systems, like mixtures of noble
gases. For instance, in Ref. \cite{74,74a} a minimum of $c_s$ in a
mixture of argon and helium at high pressure was reported. The
same authors also performed measurements of a speed of sound in
mixtures of some other gases at high pressure (carbon dioxide -
helium, nitrogen - methane and some others \cite{75,76,77,78}) and
found minima of $c_s$ as a function of concentration of the
components in some of these mixtures (carbon dioxide - helium),
but not in others (nitrogen - methane). Moreover, in the latter
case it was found that the speed of sound in a mixture at high
pressure can be relatively precisely described by an expression
for the speed of sound in a mixture of ideal gases \cite{78}. At
the same time the authors do not give any explanations of the
nature of different behavior of a speed of sound in mixtures of
different gases.

The goal of the present paper is to find the reasons of minima of
the concentration dependence of a speed of sound in mixtures of
noble gases at high pressure.

\section{System and Methods}

In the present work we simulate a mixture of argon and helium at
three different values of pressure: 200, 300 and 400 MPa at
$T=298$ K by means of the molecular dynamics method. These
thermodynamic conditions are taken from Ref. \cite{74} in order to
compare the results of simulation with experimental data. The
molar concentration of helium is varied from $x=0.0$ (pure argon)
to $x=1.0$ (pure helium) with step $\Delta x=0.1$. In all cases a
system of 32000 particles with a given concentration of species
placed in a cubic box with periodic boundary conditions is
considered. We simulate the system under constant pressure
conditions, i.e. constant number of particles N, pressure P and
temperature T. The time step is set to 1 fs. The equilibration
period is 200 ns and further 300 ns are used for calculation of
the properties of the system.

We calculate the adiabatic speed of sound as $c_s=\left( \gamma
\frac{\partial P}{\partial \rho} \right)^{1/2}$, where
$\gamma=c_p/c_V$. In order to evaluate the pressure derivative on
density we calculate two points with slightly higher density and
two more at slightly smaller one from the minimum. The pressure at
these five points is fitted to a straight line, the slope of which
gives the derivative. Analogously, by calculating two points at
higher and lower temperatures either along the isochor, or along
the isobar we calculate the heat capacities at a constant volume
or at constant pressure.

Both argon and helium were simulated with the Lennard-Jones (LJ)
potentials. The potential parameters for both helium and argon are
taken from Ref. \cite{ljparam}: $\varepsilon_{He} =10.9 $ K,
$\sigma_{He}=2.64$ $\AA$ and $\varepsilon_{Ar}=119.8$ K,
$\sigma_{Ar}=3.405$ $\AA$. The cross-interaction parameters were
taken from the Lorentz-Berthelot rules.

All simulations were performed using the LAMMPS simulation package
\cite{lammps}.

\section{Results and Discussion}

The speed of sound of the argon - helium mixture at $P=200$ MPa is
shown in Fig. 1. We see that $c_s$ from simulation
demonstrates a minimum, but it is less pronounced compared with
the experimental curve. It is seen that the speed of sound of pure
argon from simulation is in perfect agreement with the
experimental one ($c_s=1012.52$ $m/s$ from MD and $c_s=1019.64$
$m/s$ from the experiment), while the speed of sound of pure
helium in molecular dynamics is higher than the experimental one.
The speed of sound of the mixture obtained in simulation is also
higher than the experimental one at the same concentration, which
should be related to overestimation of the speed of sound of
helium. As a result, the minimum of $c_s$ from the simulation is
less pronounced than that from the experiment. However, the
location of the minimum at concentration $x_{He} \approx 0.3$ is
the same for both experiment and simulation.

\begin{figure}

\includegraphics[width=8cm, height=6cm]{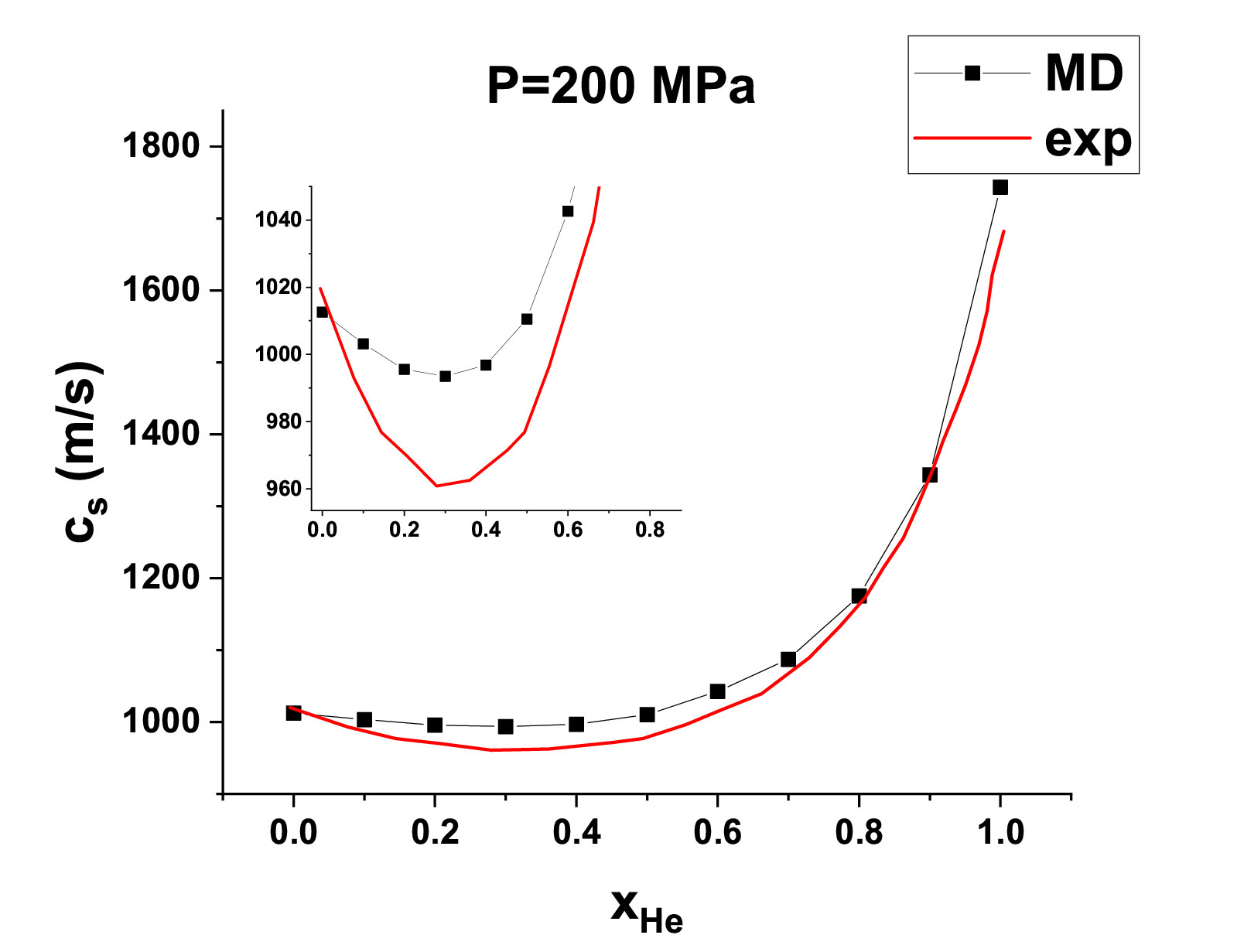}%

\caption{\label{cs200} The speed of sound in the mixture of argon
and helium as a function of helium concentration at $P=200$ MPa.
The experimental data are taken from Ref. \cite{74}. The inset
enlarges the region near the minimum of the curves.}
\end{figure}

Figures 2 (a) and (b) show the heat capacities and the
density derivative of the pressure of the argon-helium mixture at
$P=200$ MPa. It is seen that both isobaric and isochoric heat
capacities monotonically decrease with an increase in the helium
concentration. At the same time the derivative $\left(
\frac{\partial P}{\partial \rho} \right)_T$ demonstrates a minimum
at concentration $x_{He} \approx 0.2$, which is smaller than the
concentration for the minimum of $c_s$. It means that the minimum
is related to the shape of the derivative of compressibility,
while the heat capacity only shifts the minimum, but does not
change the qualitative shape of the curve.

\begin{figure}

\includegraphics[width=8cm, height=6cm]{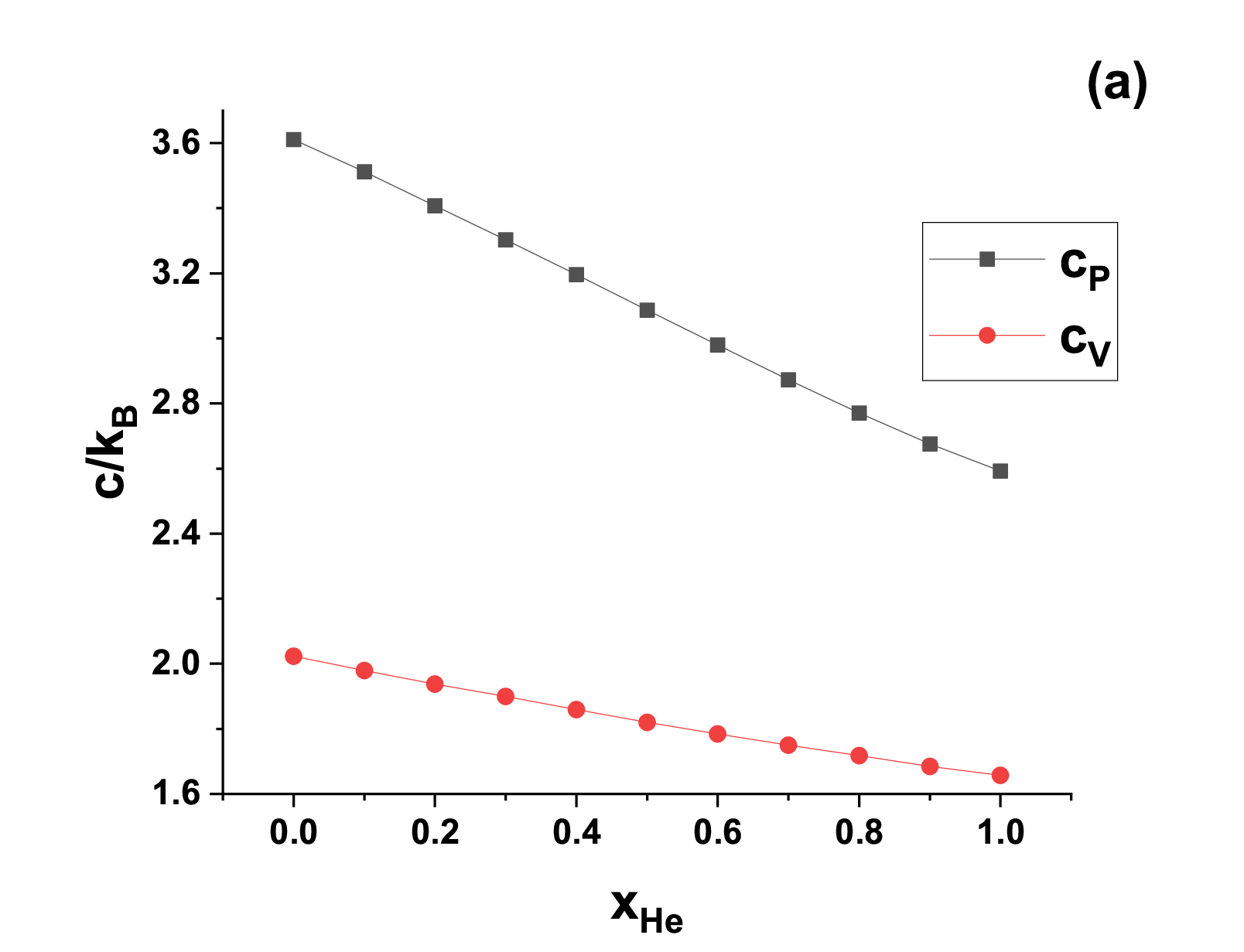}%

\includegraphics[width=8cm, height=6cm]{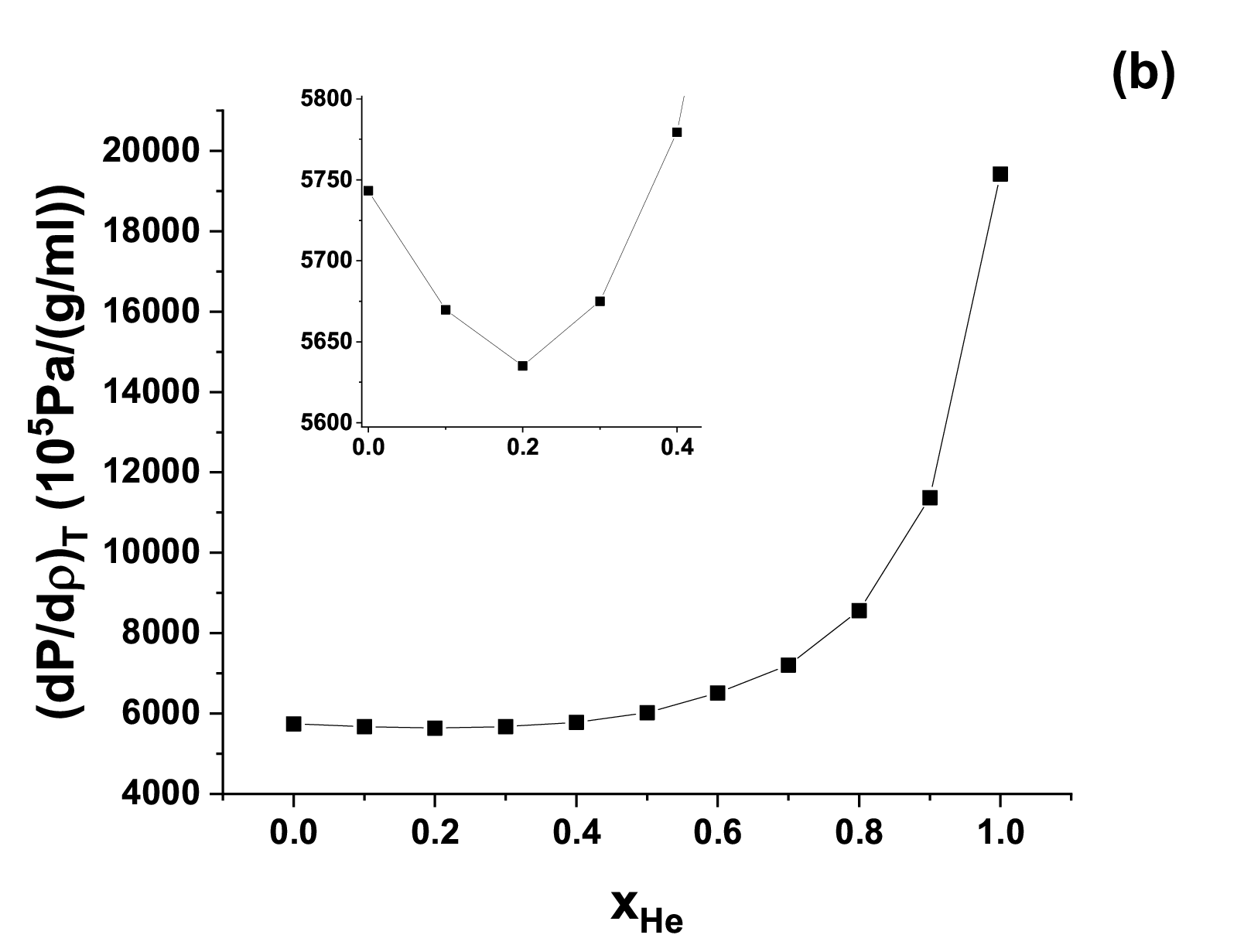}%

\caption{\label{cp200} (a) The isobaric and isochoric heat
capacities of the mixture of argon and helium as a function of
helium concentration at $P=200$ MPa (in the units of $k_B$). (b)
The density derivative of pressure of the same system. The inset
enlarges the region of the minimum.}
\end{figure}

Figure 3 shows the speed of sound in the Ar-He
mixture at pressures 300 and 400 MPa. The experimental results are
qualitatively similar to the case of 200 MPa, while the minimum of
$c_s$ from the simulation becomes less pronounced and is hardly
visible at $P=400$ MPa.

\begin{figure}

\includegraphics[width=8cm, height=6cm]{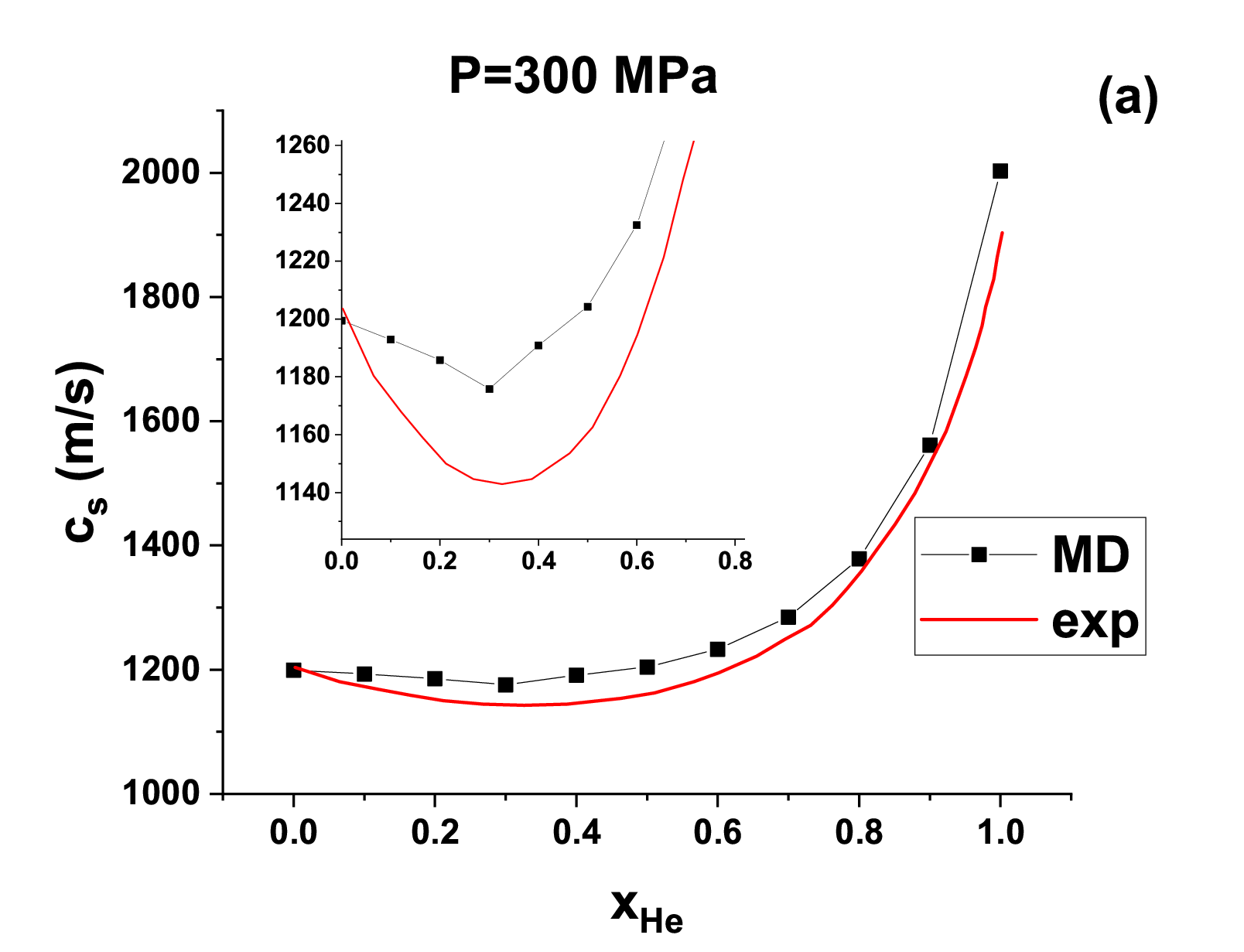}%

\includegraphics[width=8cm, height=6cm]{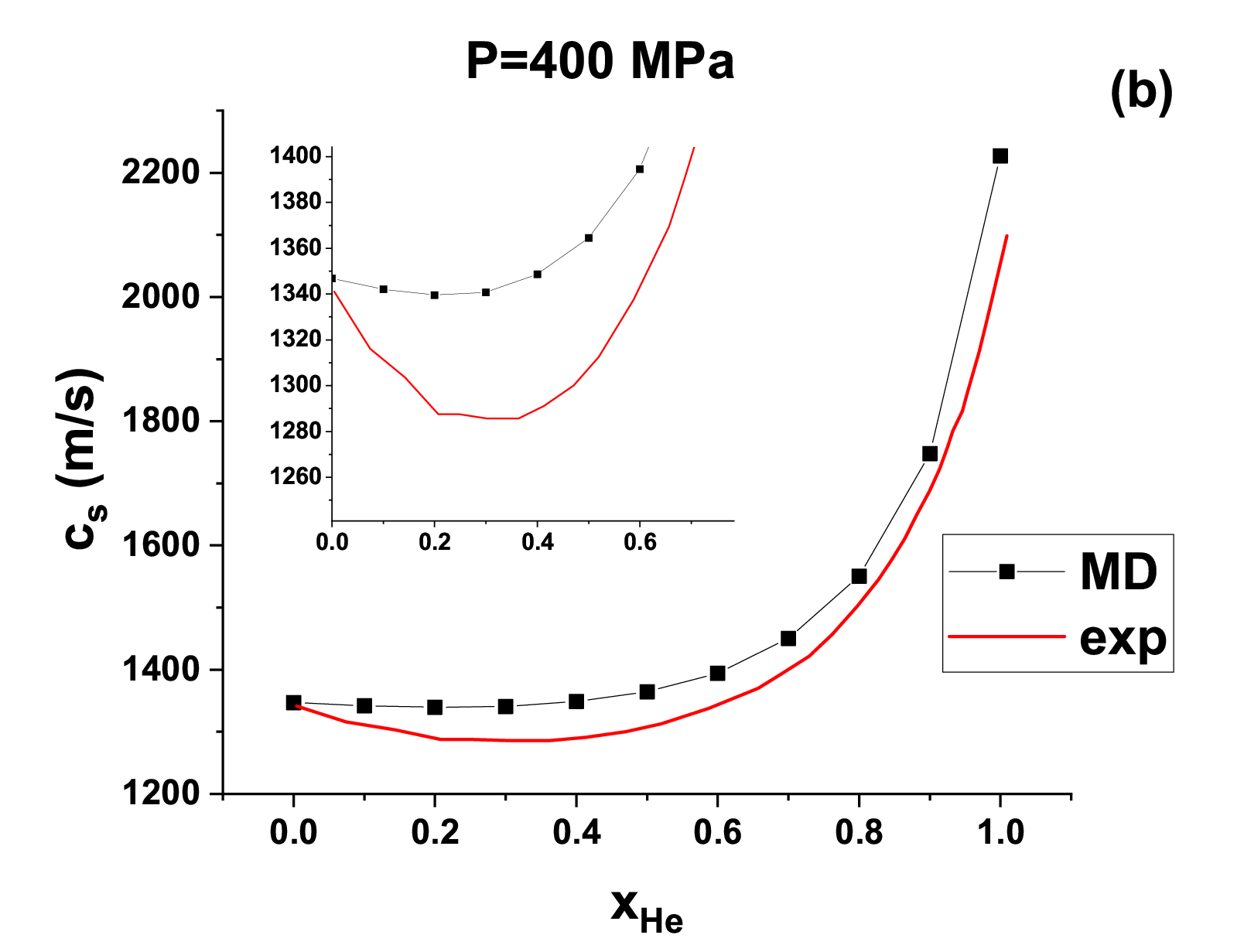}%

\caption{\label{cs300400} The speed of sound in the mixture of
argon and helium as a function of helium concentration at (a)
$P=300$ MPa and (b) $P=400$ MPa. The experimental data are taken
from Ref. \cite{74}. The insets enlarge the region near the
minimum of the curves.}
\end{figure}

%Figure \ref{pd} shows phase diagram of argon and helium placed on the same plane. The phase diagram is obtained from the LJ parameters for both gases.
%Typically a minimum of speed of sound is observed in the vicinity of the gas-liquid critcal point. From Fig. \ref{pd} it is seen that the location of minima of $c_s$
%is far from the critical points of both components. Alhougth, critical temperature of a mixture can exceed the one of both pure components (gas-gas transition),
%the minima of speed of sound take place at larger pressures, so the influence of the critical point is highly unlikely. At the same time the points of minima of
%the speed of sound are close to the Frenkel line of pure argon. Although Frenkel line is widely studied both theoretically and eperimentally for many fluids, there is a
%lack of studies of the Frenkel line of mixtures. It  looks very likely that minimum of $c_s$ can be related to the Frenkel line of argon, which, however, requires further investigations.

%\begin{figure}

%\includegraphics[width=8cm, height=6cm]{arhe-pd.eps}%

%\caption{\label{pd} Phase diagram of models of Ar and He employed in the present paper. Frenkel lines and location of minima of speed of sound of the mixtures are added.
%The melting line is taken from Ref. \cite{lj-melting} and the boiling line from Ref. \cite{lj-boiling}. The lines denotes as 'Frenkel $c_V=2$' and $Frenkel vacf$ are the Frenkel lines
%calculated via isochoric heat capacity and velocity autocorrelation functions respectively.}
%\end{figure}

It is known that mixtures can demonstrate much more complex phase
diagrams than pure substances (see, for instance, \cite{row}). In
particular, a mixture of argon and helium belongs to the III type
in the van Konynburg - Scott classification \cite{kon}, i.e. in
addition to a usual gas - liquid transition it also demonstrates a
liquid-liquid one at high pressures. There are several works where
the phase diagram of an Ar-He mixture is reported
\cite{arhe-pd-1,arhe1,arhe2}. An empirical equation of state for
this mixture is constructed in Ref. \cite{arhe-pd-refs} and the
phase diagram of the mixture is calculated. However, all these
data belong to either much lower pressures, or much lower
temperatures. To the best of our knowledge, no data for phase
behavior of an Ar-He mixture at $T=300$ K and pressures above
$200$ MPa are reported in the literature. Figure 4(d) of Ref.
\cite{arhe-pd-refs} reports the phase diagram of the mixture up to
very high pressure $P=10$ kbar and temperature up to $180$ K.
Extrapolation of the data of this figure leads us to the
conclusion that the points of interest ($T=300$ K and $P=200$,
$300$ and $400$ MPa) belong to one phase region, and should be far
from the phase transition lines. Therefore, no influence of phase
transitions is expected at these points.

It is well known that a speed of sound shows a minimum along
isobars. Sometimes this minimum is considered as a boundary of
liquid-like (the speed of sound decreases with temperature) and
gas-like (the speed of sound increases with temperature) regimes
in supercritical fluid. Interestingly, the minimum of the speed of
sound is located close to the Frenkel line of the fluid (although
not exactly on the Frenkel line). This minimum of the speed of
sound takes place at relatively high temperature. For instance, in
the case of argon at $P=200$ MPa the speed of sound demonstrates a
minimum at $T_m=681$ K \cite{nist}. The minimum of the speed of
sound of helium at the same pressure corresponds to $T_m=188.5$ K
\cite{nist}. The phase diagrams of argon and helium are very close
to the phase diagram of the LJ system \footnote{The quantum effect
for helium is negligible in the range of thermodynamic parameters
of this work}. Figure 4 shows the phase diagram of the
LJ system in the units of $P/P_c$ and $T/T_c$. We show the Frenkel
line and the locations of minima of the speed of sound in the
mixture obtained in the present work. For the sake of comparison,
we also show the location of minima of $c_s$ along the isobars of
pure argon and pure helium taken from the NIST database
\cite{nist}. One can see that at large $P/P_c$ the line of minima
of $c_s$ becomes roughly parallel to the Frenkel line in the
double logarithmic coordinates. At the same time for pure
substances, it is always slightly above the Frenkel line.

%As it was shown before, Frenkel line starts at the boiling line at the temperature about $0.8T_c$, where $T_c$ is the critical point \cite{ufn-fr,pre-fr,jept-fr,scf-fr}. Therefore,
%as lower the critical point, as lower the Frenkel line. The critical line of the mixtrure of argon and helium was measured in Ref. \cite{arhe-pd-1}.
%It was shown that the critical temperature demonstrates a minimum as a function of the concentration of helium at the concentrations
%$c_{He} \approx 0.5$. The minima obtained in the present study correspond to $c_{He}=0.3$, so the critical temperature of the
%gas-liquid transition of this mixture is lower than the one of pure argon. As a result the Frenkel temperature of the mixture at given pressure is
%lower than the one of pure argon at the same pressure. The minimum of the speed of sound should also shift to lower temperatures.

The point with given thermodynamic variables (P,T) corresponds to
different points for argon and helium in the plane
($P/P_c$,$T/T_c$), since these gases have strongly different
critical parameters. The points corresponding to the mixtures
should belong to a line connecting those of pure components at the
same pressure. As is seen from Fig. 4 this line (the
dotted line) should be located not far from the Frenkel line of
the LJ system and even cross it. Thus, trajectories corresponding
to the minima of the speed of sound under the change of the
concentration of the components near the border between gas-like
and liquid-like regimes of fluid, appear on the phase diagram of
the LJ system. One should expect similar behavior for the other
properties of fluid demonstrating minima at the isobars (shear
viscosity, heat conductivity, etc.).

%The corresponding critical temperatures
%are $T_c=150.7$ K for argon and $T_c=5.2$ for helium respectively. Effective one can consider that changing the concentration
%of the components from pure heavy one to the pure light one shifts the system on the phase diagram from the points denoted as 'Ar'
%to the ones denotes as 'He' in Fig. \ref{pd-crit}.

As it was discussed in a number of works
\cite{ufn-fr,scf-fr,fr-solub}, the properties of liquid in the
vicinity of the Frenkel line can be optimal for some technological
applications. At the same time the thermodynamic parameters of the
Frenkel line can be hardly achievable for many industrial fluids,
since they involve rather high pressures. From the results of the
present paper, we can conclude that the Frenkel line of a
particular fluid can be shifted by mixing it with light components
with a low critical point. This conclusion opens a window for
novel methods of design of fluids with a given position of the
Frenkel line on the phase diagram. It is also desirable to perform
investigation of concentration dependence of other properties of
fluid mixtures in the vicinity of the Frenkel line.

%We are not aware of any data on the critical point of mixture of argon and helium. One might expect from the parameters
%of LJ interaction used in the present work that this mixture should demonstrate type III behavior in van Konynburg-Scott classification.
%It means that in addition to the gas- liquid phase transition a liquid-liquid one at high pressure appears in the system. However,
%we do not observe any discontinuties in the behavior of the system. For this reason we believe that no phase transition lines are
%crossed in our simulations. However, the changing the concentration of the components effectively shifts the system from
%the points denoted as 'Ar' in Fig. \ref{pd-crit} to the ones denoted as 'He'. It can lead to cross of the Frenkel line, which is
%near to the points of minima of the speed of sound. It means that changing the concentration from pure heavy component to the pure liquid one
%effectively changes the fluid from liquid-like to the gas-like one.

\begin{figure}

\includegraphics[width=8cm, height=6cm]{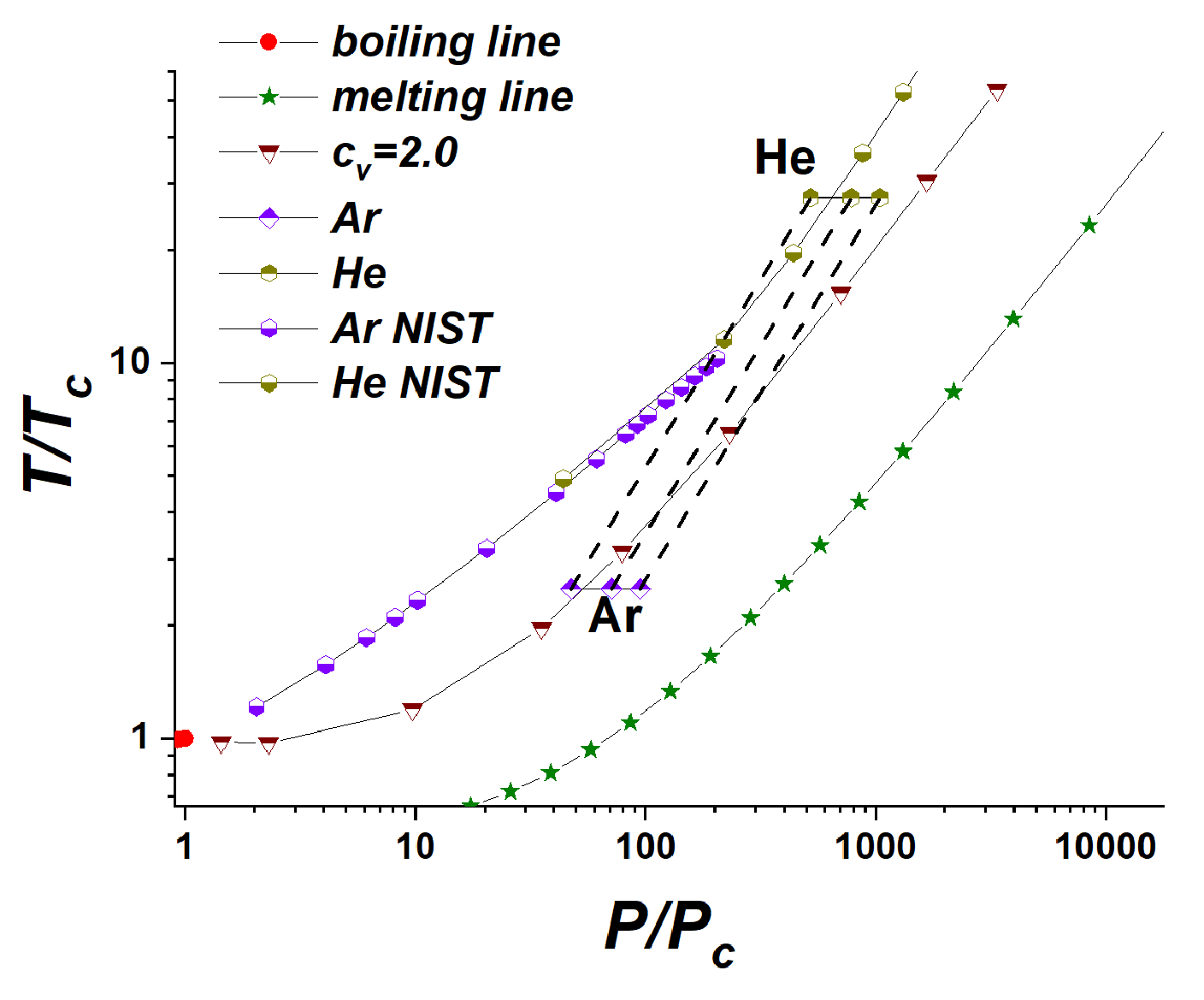}%

\caption{\label{pd-crit} The phase diagram of the LJ system in
reduced units. The points 'Ar' show the points of pure argon at
$T=300$ K and $P=200$, $300$ and $400$ MPa, i.e., the
thermodynamic conditions studied in this paper. The points denoted
as 'He' show the same for helium. Line $c_V=2.0$ shows the
location of the Frenkel line. The points 'Ar NIST' and 'He NIST'
show the location of minima of the speed of sound along the
isobars in pure argon and helium taken from the NIST database
\cite{nist}. The boiling line is taken from. Ref.
\cite{lj-boiling} and the melting line from Ref.
\cite{lj-melting}.}
\end{figure}

\section{Conclusions}

A minimum of the speed of sound in a mixture of the noble gases
argon and helium was established experimentally a long time ago.
In the present paper we perform a molecular dynamics study of this
system based on a simple LJ model of the gases. We show that this
simple model also demonstrates a minimum of the speed of sound as
a function of the concentration of helium. The location of the
minimum is in good agreement with the experiment. We assume that
this minimum might be related to the proximity to the Frenkel line
of the mixture.

This work was carried out using the computing resources of the
federal collective usage center "The complex for simulation and
data processing for mega-science facilities" at NRC "The Kurchatov
Institute", http://ckp.nrcki.ru, and the supercomputers at the
Joint Supercomputer Center of the Russian Academy of Sciences
(JSCC RAS). The work was supported by the Russian Science
Foundation (Grant 24-12-00037).

\end{document}